\documentclass[reprint,aps,pof]{revtex4-1}

\usepackage{graphicx}

\begin{document}

\title{The Hama Problem revisited: essential mixing in a free shear flow}
\author{V. A. Miller}
\email[Corresponding author: ]{vamiller@stanford.edu}
\affiliation{Stanford University}

\author{M. G. Mungal}
\affiliation{Stanford University}
\affiliation{Santa Clara University}

\begin{abstract}
We revisit a problem first introduced by Francis Hama in his 1962 \textit{Physics of Fluids} article \textquotedblleft Streaklines in a Perturbed Shear Flow." Using a nascent computer, Hama calculated streaklines and pathlines for an inviscid shear flow containing  a non-amplifying, sinusoidal, traveling-wave perturbation.  He found that this simple flow field and perturbation produced intuitive pathlines, yet, the resulting streaklines were non-obvious and much more complex than the pathlines.  He used this work to stress the importance of understanding the difference between flow visualization and measurement techniques, and that different techniques may reveal different information regarding the true character or behavior of the flow.  This work revisits the Hama Problem and presents the original findings in a fluid dynamics video prepared for the 2013 Gallery of Fluid Motion.\end{abstract}

\maketitle

\section{Introduction}
A variety of techniques have enabled researchers to study and understand structure, turbulence, mixing, and the nuanced physics of fluid dynamics: hot wire anemometry can be used to sensitively and accurately measure fluctuating velocities; powerful PIV techniques can record time-resolved, 3D velocity fields; smoke wires, streams of dye can reveal fluid particle paths, streaks, and the scalar mixing field; tufts of string and oil slicks can inform us about boundary layer behavior; and sophisticated simulations are used to predict our observations or model the unobservable.  

This work revisits a flow configuration raised by Hama \cite{Hama1962}, a free shear flow with a non-amplifying, small amplitude, traveling-wave perturbation in $u$ and $v$.  Hama computed the pathlines and streaklines for the flow, and he discovered a striking difference between the two.  Neither form of flow visualization (i.e. pathlines and streaklines) is at all informative of the other, which raises questions as to what exactly the relevant nature or character of a flow field is, and how to best go about probing, measuring, or understanding that character.  Building on the seminal work, we also compute timelines in addition to pathlines and streaklines, and we present the results in this manuscript as well as a video.

\section{The perturbed shear flow}
\begin{figure}
\begin{centering}
\includegraphics[width=6.5cm]{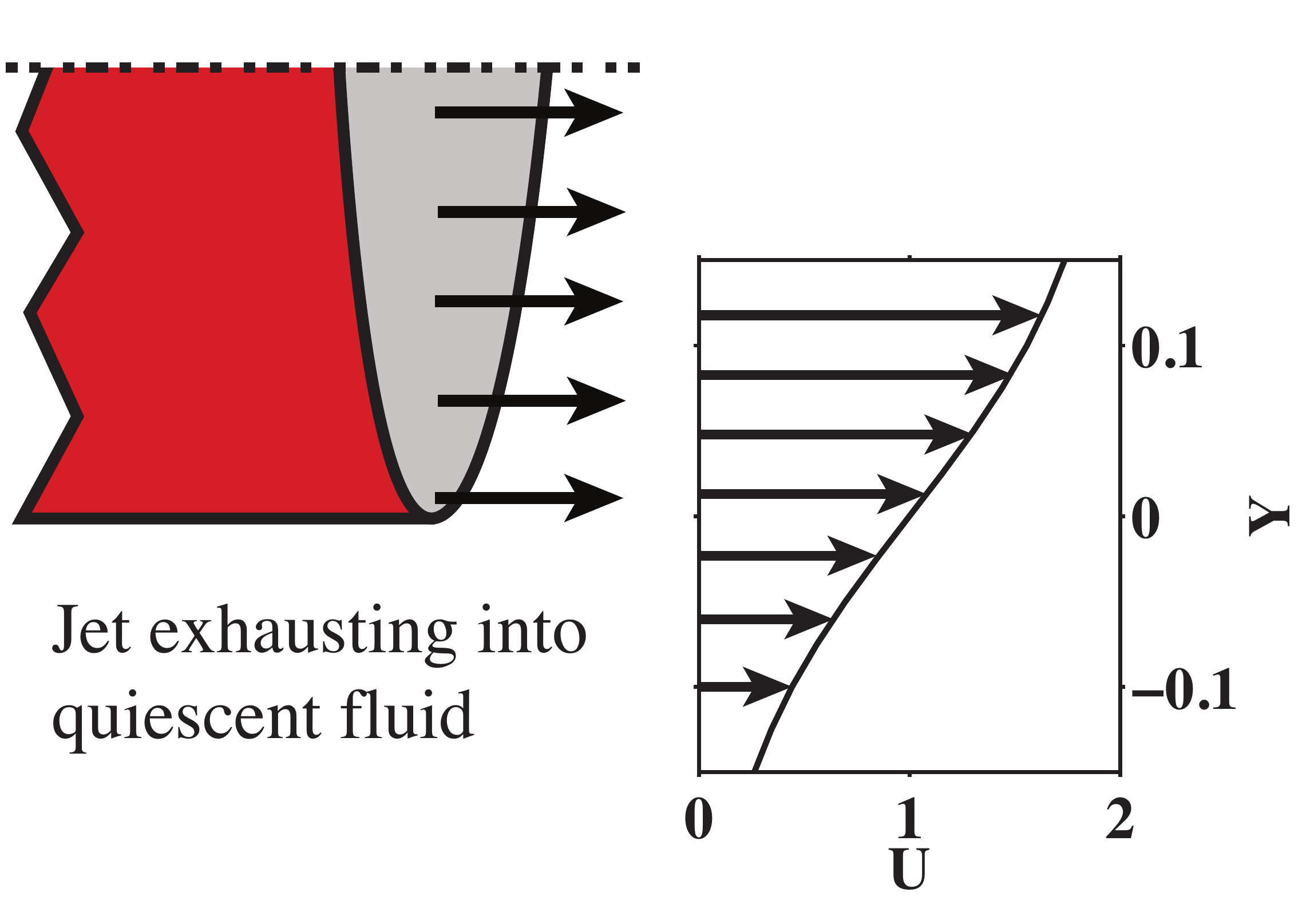}
\par\end{centering}
\caption{\label{fig:unpertVel}Schematic of flow field of interest.  This flow field could be found, for example, in a jet exhausting in quiescent fluid.}
\end{figure}

\begin{figure}
\begin{centering}
\includegraphics[width=9cm]{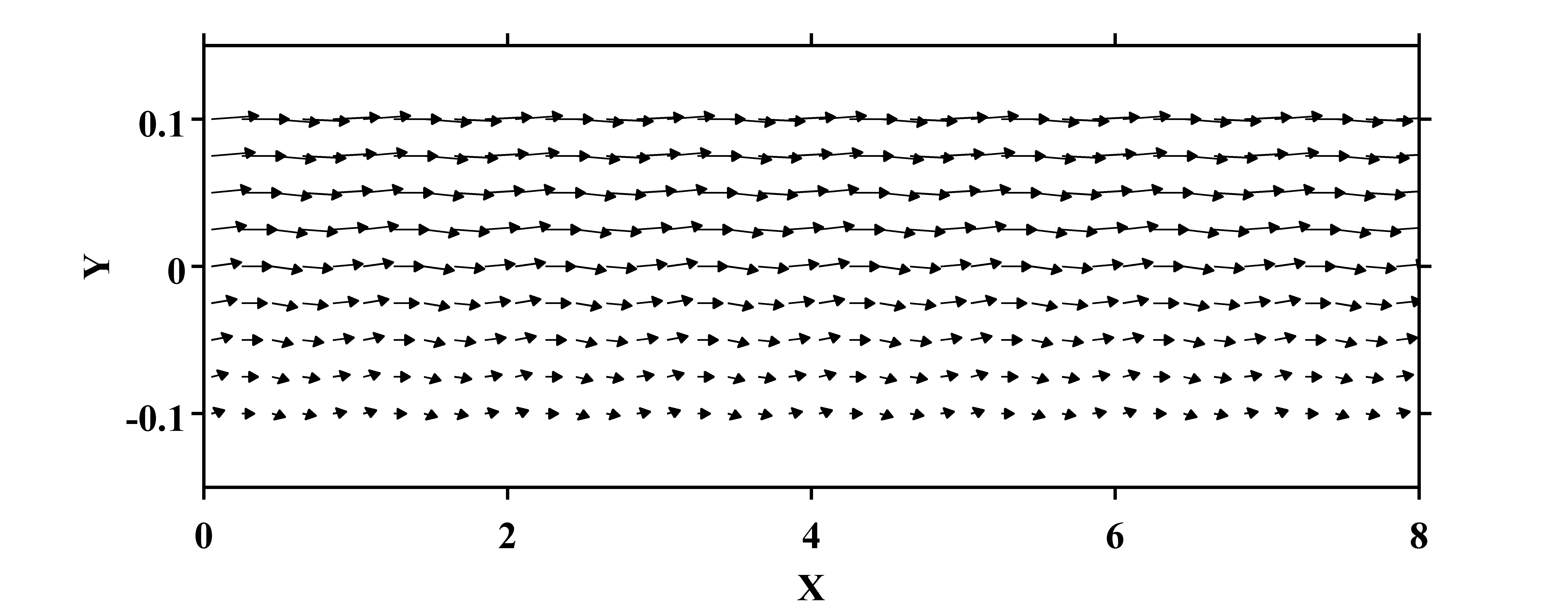}
\par\end{centering}
\caption{\label{fig:petVel}Vector field for perturbed vector field.}
\end{figure}

The basic, unperturbed flow for this work is described by Eq. 1.
\begin{equation}
u_o=1+tanh(y), \qquad v_o=0
\end{equation}
This velocity field could be found in, for example, a jet flowing into quiescent fluid (Fig. \ref{fig:unpertVel}).  
The perturbations in $u$ and $v$ are given by Eq. 2, 

\begin{eqnarray}
u' &= &2a \text{ sech}(y) \tanh(y) \sin(\alpha(x-ct)) \\
v' &= &2a \text{ sech}(y) \cos(\alpha(x-ct)) \nonumber
\end{eqnarray}

 where $\alpha$ and $c$, the wavenumber and wave velocity of the perturbation, are, respectively unity.  $a$ is the perturbation amplitude, and we have chosen $a=0.015$ for this work.  For reference, in the original work \cite{Hama1962}, $a=$ $0.005$, $0.01$, and $0.02$ were investigated.  The perturbation is a traveling-wave, where $v'$ is symmetric about $y=0$ depending on $\text{sech}(y)$, and $u'$ is the product of $\text{sech}(y)$ and $\tanh(y)$.   For ease of computation and display of the results, we transform the coordinate frame of the velocity equations via 
 \begin{eqnarray}
 X&=&x/\lambda, \qquad Y=y/\lambda, \nonumber \\
 T&=&ct/\lambda, \qquad \lambda=2\pi, \nonumber
 \end{eqnarray}
 and therefore, axes in this work are labeled with uppercase variable names.  Details of the transformation can be found in \cite{Hama1962}.  Fluid particle paths are solved for using $MATLAB$ and its ordinary differential equation solver, $\texttt{ode45()}$.  The perturbed velocity field at an instance in time is presented in Figure \ref{fig:petVel}.

\section{Pathlines, streaklines, and timelines}
\begin{figure}
\begin{centering}
\includegraphics[width=9cm]{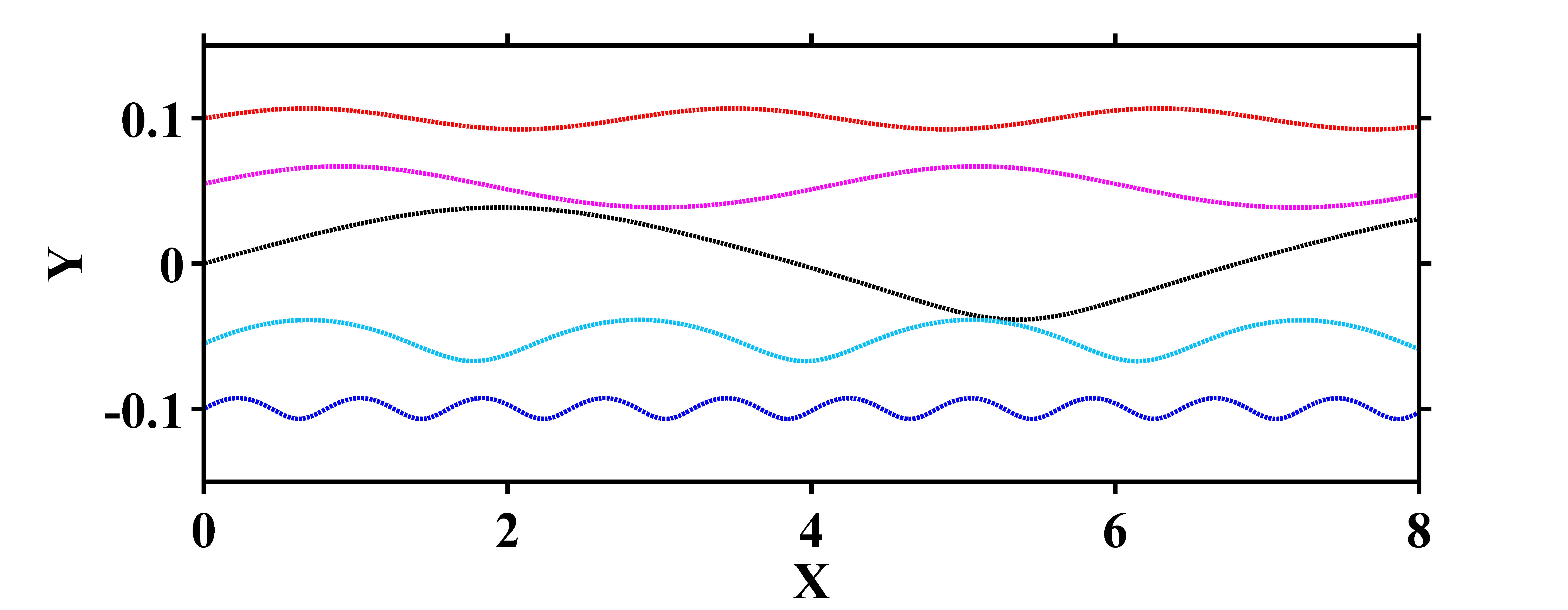}
\par\end{centering}
\caption{\label{fig:paths}Pathlines positioned at increments of $\Delta Y=.05$ centered about $Y=0$.  Given the perturbed velocity field, pathlines are generally intuitive, suggesting a sinusoidal, traveling-wave perturbation.  Pathlines presented are all released at $T=0$.}
\end{figure}

We first plot the pathlines of the flow (Fig. \ref{fig:paths}); the pathlines presented are for fluid particles that are all released at $T=0$.  Individual particles travel along sinusoidal-like paths, and the wavelength of each particle's path depends upon its initial $Y$ location.  Particles in the faster flow near the top of the domain travel downstream in less time than those at the bottom.  Particle paths for particles near the top of the domain have a longer wavelength, because their convective velocity is larger relative to the perturbation wave velocity $c$, and, as expected, paths of particles near the bottom of the domain have shorter wavelengths.  Particles released at other times follow pathlines of similar character to the ones shown here, that is, simple sinusoidal-like paths for which wavelength depends on the starting location of the particle.  The pathlines are intuitive, yet, from them, the fluctuations in $u$ and $v$ observable in Figure \ref{fig:petVel} are not immediately apparent.

\begin{figure*}
\begin{centering}
\includegraphics[width=16cm]{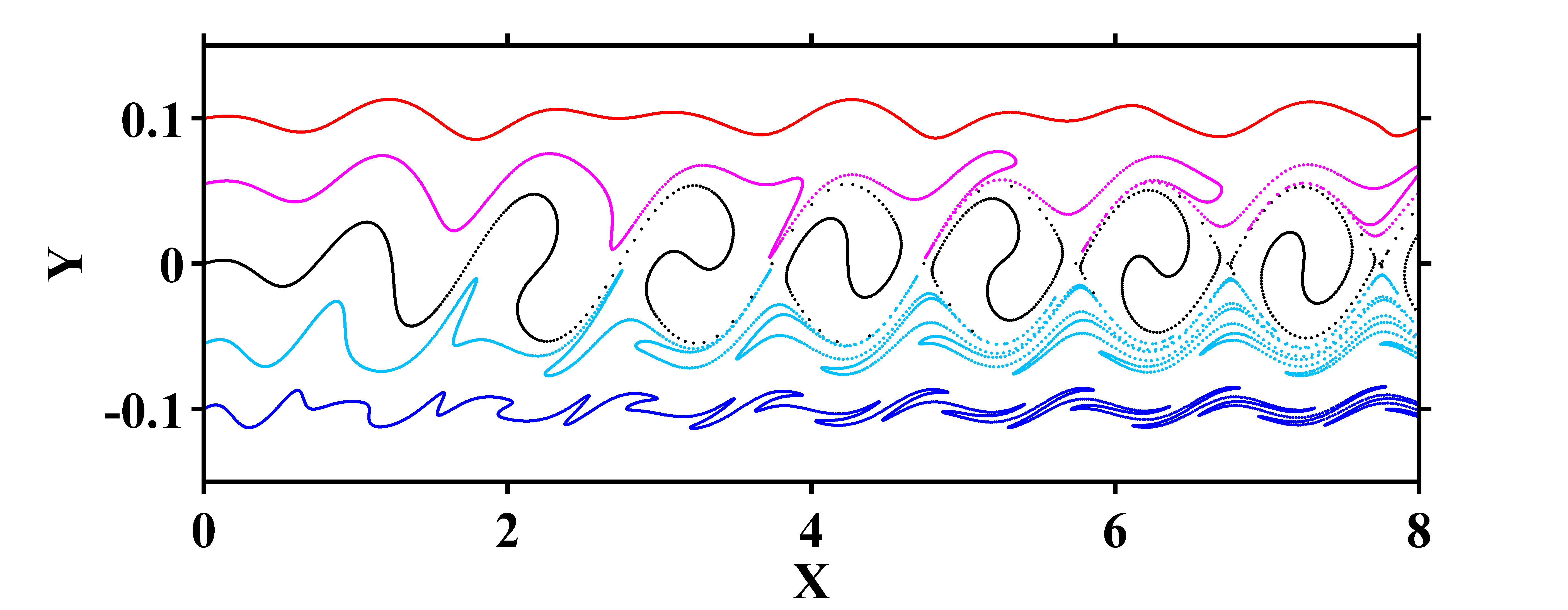}
\par\end{centering}
\caption{\label{fig:singleStreaks}Streaklines at $T=20$ originating at the same locations as the pathlines in Fig. \ref{fig:paths}.}
\end{figure*}

For the same particle starting locations, streaklines at $T=20$ are shown in in Figure \ref{fig:singleStreaks}.  Streaklines are created by tracking individual particles of \textquoteleft numerical dye,' released at a constant rate from $X=0$.  The streaklines are starkly different from the pathlines.  Initially, near $X=0$, the streaklines are what we expect, i.e., sinusoidal patterns, but only a short distance downstream (i.e., around $X=2$), the streaks begin rolling up into \textquoteleft cat's-eye' structures.  Streaks originating above and below a certain $Y$ value (roughly, between $Y=\pm0.05$ for this case, referred to as the \textquotedblleft critical layer" in \cite{Hama1962}) are rolled into these structures.  Above $Y\approx0.05$, particles convect downstream, leaving wavy streaklines, and below $Y\approx-.05$, the streaklines form their own, complicated, folded structure.  Despite the streaklines rolling up into the cat's-eye structures, no discrete vortices exist in the flow.  Further downstream ($X>3$), we also observe a spreading out of dye particles as the filaments are stretched.

\begin{figure*}
\begin{centering}
\includegraphics[width=16cm]{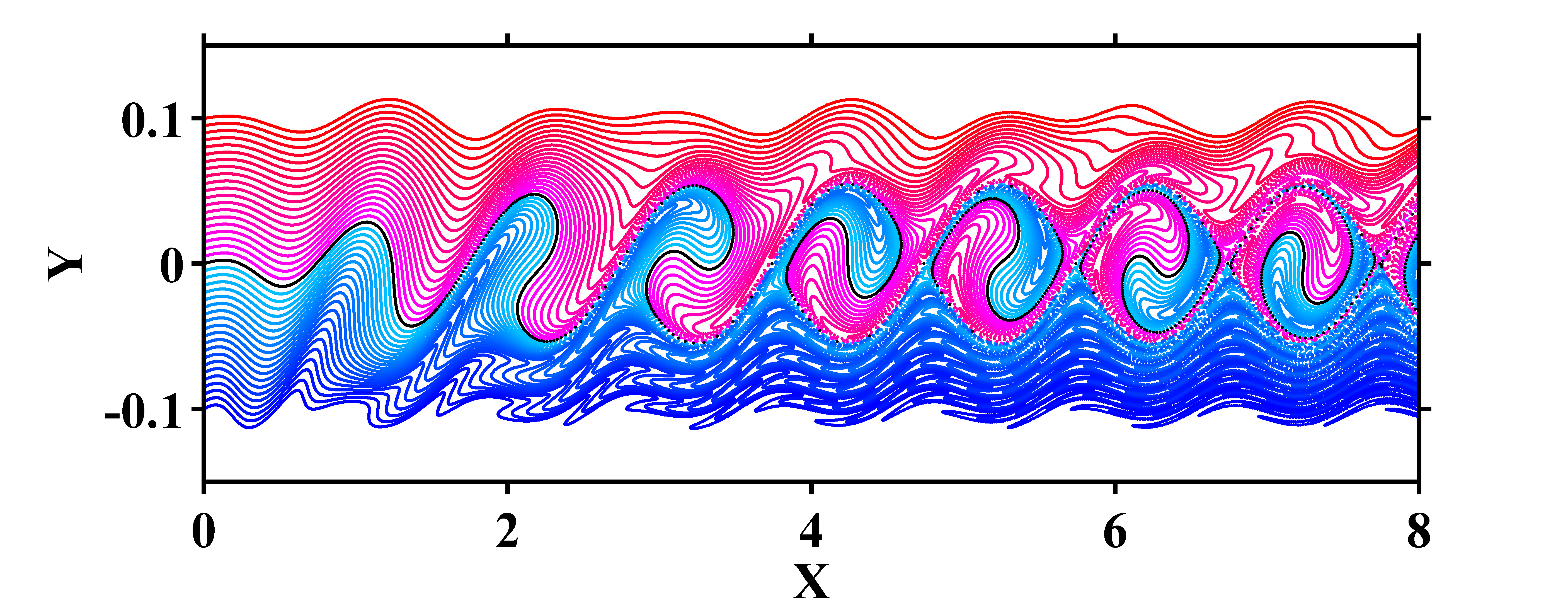}
\par\end{centering}
\caption{\label{fig:FullStreaks}Streaklines at $T=20$ positioned at increments of $\Delta Y=.005$ centered about $Y=0$.}
\end{figure*}

Streaklines generated at intervals of $\Delta Y = 0.005$ are shown in Figure \ref{fig:FullStreaks}, which highlights the complexity of the scalar mixing field, which has resulted from a simple analytic velocity field.  

\begin{figure*}
\begin{centering}
\includegraphics[width=16cm]{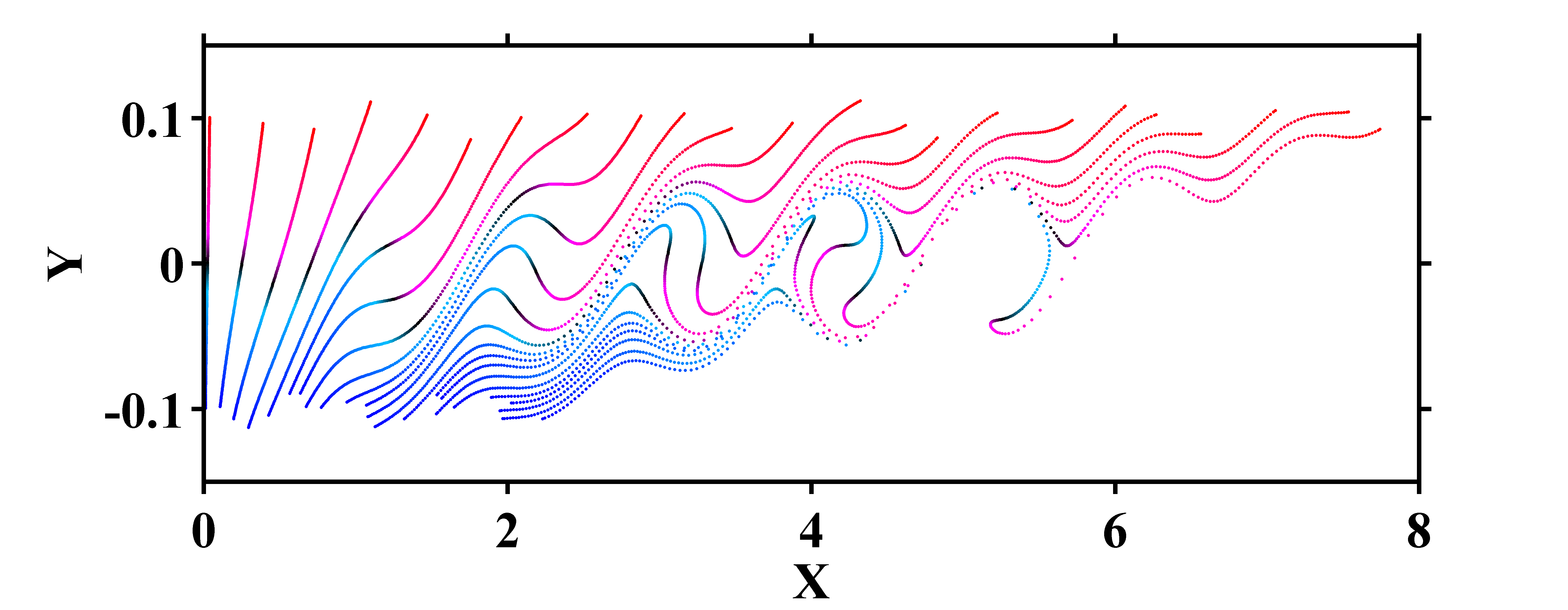}
\par\end{centering}
\caption{\label{fig:timelines}Timelines at $T=5.00$.  Timelines are released every $\Delta T=0.225$.}
\end{figure*}

Lastly, we present timelines in Figure \ref{fig:timelines}, where a vertical band of dye is released from $X=0$, spanning from $Y$ from $-0.1$ to $0.1$, and timelines are released every $0.225$ units of time.  Again, this vertical band of numerical dye is created by tracking individual particles of dye; each timeline contains $2000$ particles of dye, and again, we can see that as each timeline is stretched throughout the domain, the density of dye decreases, especially around the perimeter of the cat's-eye structures.  The first timeline released is stretched from about $X=2$ all the way to $X=8$, circumnavigating nearly an entire cat's-eye structure.  Other timelines indicate that depending upon when the timeline is released, its middle section (the black region) can either get caught inside the cat's-eye, or it can be stretched out between the rollers.  

\section{Conclusion}
In this brief manuscript, we revisit the original work of Francis Hama, computing the pathlines, streaklines, and timelines of a simple shear flow with a traveling-wave perturbation.  \textquotedblleft In conclusion...practically no truth can be obtained from the streakline or pathline observations as to the nature of time-dependent phenomena." \cite{Hama1962}.  The vector field, pathlines, streaklines, and timelines all reveal very different information regarding the fluctuating nature of the flow field and scalar mixing field.

\begin{acknowledgments}
V. A. Miller is supported by the William and Claudia Coleman Stanford Graduate Fellowship.\end{acknowledgments}

\bibliographystyle{abbrv}
\bibliography{library.bib}

\end{document}